\documentstyle[12pt]{article}
\newcommand{\Z}{{\sf Z \!\!\! Z}}

\newcommand{\1}{{\sf 1 \!\! 1}}
\newcommand{\Sig}{{\sf \Sigma \!\!\! \Sigma}}
\makeatletter
\@addtoreset{equation}{section}
\makeatother
\title{Bosonization and Cluster Updating of Lattice Fermions}
\author{U.-J. Wiese$^+$ \\[2em]
Universit\"at Bern, Sidlerstrasse 5, 3012 Bern, Switzerland}
\begin{document}

\maketitle

\begin{abstract} \normalsize

A lattice fermion model is formulated in Fock space using the Jordan-Wigner
representation for the fermion creation and annihilation operators. The
resulting path integral is a sum over configurations of lattice site
occupation numbers $n(x,t) = 0,1$ which may be viewed as bosonic Ising-like
variables. However, as a remnant of Fermi statistics a nonlocal sign factor
arises for each configuration. When this factor is included in measured
observables the bosonic occupation numbers interact locally, and
one can use efficient cluster algorithms to update the bosonized variables.

\end{abstract}

\vspace{5cm}
\begin{flushleft}
$^+$ supported by Schweizer Nationalfond
\end{flushleft}
\pagebreak

Numerical simulations of lattice fermions are important both in particle and
in condensed matter physics. For example, in particle physics one simulates
quarks to investigate if nonperturbative QCD is the correct theory of strong
interactions. In condensed matter physics one simulates electrons to
test if the Hubbard model and its variants
describe high $T_c$ superconductivity. In both
cases it is common to integrate out the fermions. The resulting bosonic
effective theories have nonlocal interactions due to the fermion determinant.
Because of the nonlocality the theory is not really bosonized, and its
numerical simulation is computationally difficult and very time consuming.
Of course, one could include the fermion determinant in measured
observables and view the system as a local bosonic theory. In practice,
however, this does not work because the fermion determinant varies over many
orders of magnitude thus making a numerical simulation extremely inefficient.

Jordan and Wigner were the first to realize that fermions can be
represented by bosonic operators \cite{Jor28}. Their observation was
applied to numerical simulations of fermionic systems in
$1+1$ dimensions \cite{Hir82}. The resulting path integral is a sum over
configurations of occupation numbers $n(x,t) = 0,1$
which may be viewed as bosonic Ising-like variables.
When generalized to higher dimensions unpleasant minus-signs arise which
may cause problems in numerical simulations. Still, Duncan \cite{Dun88} could
simulate a model of staggered fermions in $2+1$ dimensions.
Later Montvay \cite{Mon89} refined the method and applied it to
Wilson fermions in $3+1$ dimensions. Here I present a new construction which
eliminates several constraints present in the
configurations of the previous approaches.
As before a nonlocal sign factor arises for the Boltzmann weight of each
configuration reflecting the Fermi statistics of the original theory.
Hence, the bosonic theory of occupation numbers is not fully local.
However, again one may include this factor in the measured observables
and view the system without the sign factor as a local bosonized theory.
The resulting theory corresponds to a quantum spin system.
Of course, also the sign factor may fluctuate but, as opposed to the usual
fermion determinant, the fluctuations are restricted to $\pm 1$. Still,
for some models this may lead to a
minus-sign problem of large cancellations between configurations with
positive and negative Boltzmann weights.
In previous approaches to the problem
the system of occupation numbers was updated using
standard local Metropolis or heat bath algorithms. These algorithms are known
to suffer from critical slowing down. Recently, Evertz, Lana and Marcu
\cite{Eve92} have developed cluster algorithms for vertex models, which can
also be applied to quantum spin systems. Unlike the usual Swendsen-Wang-Wolff
clusters \cite{Swe87} the Evertz-Lana-Marcu clusters form closed loops.
Since our fermion system corresponds to
a quantum spin system a cluster algorithm can be applied also here.

To be specific I restrict myself to a simple model which, however, shows the
characteristic features of more complicated (and physically much more
interesting) fermionic systems. It is straight forward (although in some cases
perhaps tedious) to apply the same ideas to more general fermionic models. I
consider fermions living on the sites of a spatially 2-dimensional
$L \times L$ lattice with even $L$ and with periodic boundary conditions. The
characteristic difficulties of fermion updating arise because the operators
$c_x^+$ and $c_x$, which create and annihilate fermions at the lattice site
$x = (x_1,x_2) \in \Z^2$, anticommute
\begin{equation}
\{c_x^+,c_y^+\} = 0, \,\,\, \{c_x,c_y\} = 0, \,\,\,
\{c_x^+,c_y\} = \delta_{xy}.
\label{anticommutators}
\end{equation}
The difficulties are due to Fermi statistics, not due to
spin. For simplicity I therefore
consider fermions without spin and with a Hamilton operator
\begin{equation}
H = \sum_{x,i} (c_x^+ c_x + c_{x+\hat{i}}^+ c_{x+\hat{i}}
- c_x^+ c_{x+\hat{i}} - c_{x+\hat{i}}^+ c_x),
\end{equation}
where $\hat{i}$ is the unit vector in $i$-direction. The model is trivial and
can be solved in momentum space by introducing
$c_p^+ = \frac{1}{L} \sum_x \exp(ipx) c_x^+$,
$c_p = \frac{1}{L} \sum_x \exp(-ipx) c_x$,
which implies
$H = \sum_p \hat{p}^2 c_p^+ c_p$ with $\hat{p}_i = 2 \sin(p_i/2)$.
For example, in the grand canonical ensemble
the expectation value of the occupation number $n_x =
c_x^+ c_x$ of a site $x$ is given by
\begin{equation}
\langle n_x \rangle = \frac{1}{Z} \mbox{Tr} [n_x \exp(- \beta (H - \mu N))] =
\frac{1}{L^2} \sum_p \frac{1}{1 + \exp(\beta (\hat{p}^2 - \mu))},
\end{equation}
where $\beta$ is the inverse temperature, $\mu$ is the chemical potential and
$N = \sum_x n_x$ is the fermion number operator.
Let us try to obtain the same result from a numerical simulation.

To write the partition function $Z$ as a path integral we first label the
lattice sites $x = (x_1,x_2)$ for $x_i \in \{0,1,...,L-1\}$ by
$l \in \{1,2,...,L^2\}$
in some arbitrary order. A convenient choice for numerical applications is
e.g. $l = 1 + x_1 + x_2 L$.
Following Jordan and Wigner \cite{Jor28}
the fermion creation and annihilation operators are represented as
\begin{equation}
c_x^+ = (-1)^{x_1+x_2} \sigma_1^3 \sigma_2^3 ... \sigma_{l-1}^3 \sigma_l^+,
\,\,\, c_x = (-1)^{x_1+x_2} \sigma_1^3 \sigma_2^3 ... \sigma_{l-1}^3
\sigma_l^-,
\end{equation}
where $\sigma_l^i$ are Pauli matrices associated with the lattice point labeled
with $l$ and $\sigma_l^{\pm} = \frac{1}{2}(\sigma_l^1 \pm i \sigma_l^2)$. The
different signs for $x_1+x_2$ even and odd lattice points are not really
necessary but they absorb an inconvenient minus-sign in some expressions below.
I decompose the Hamiltonian into four pieces $H = H_1 + H_2 + H_3 + H_4$
\begin{equation}
H_1 = \sum_{x=(2m,n)} h_{x,1}, \,\,\, H_2 = \sum_{x=(m,2n)} h_{x,2}, \,\,\,
H_3 = \sum_{x=(2m+1,n)} h_{x,1}, \,\,\, H_4 = \sum_{x=(m,2n+1)} h_{x,2},
\end{equation}
where $h_{x,i} = c_x^+ c_x + c_{x+\hat{i}}^+ c_{x+\hat{i}}
- c_x^+ c_{x+\hat{i}} - c_{x+\hat{i}}^+ c_x$.
The individual contributions to a given $H_j$ commute with each other,
but two different $H_j$ do not commute. Using the Suzuki-Trotter formula one
writes for the grand canonical partition function
\begin{eqnarray}
Z = \mbox{Tr} \exp(- \beta (H - \mu N))
= \mbox{lim}_{M \rightarrow \infty} \mbox{Tr}
[\exp(- \epsilon \beta (H_1 - \frac{\mu}{4}N)) \nonumber \\
\times \exp(- \epsilon \beta (H_2 - \frac{\mu}{4}N))
\exp(- \epsilon \beta (H_3 - \frac{\mu}{4}N))
\exp(- \epsilon \beta (H_4 - \frac{\mu}{4}N))]^M,
\end{eqnarray}
where $\epsilon \beta = \beta/M$
is the lattice spacing in the euclidean time direction.
Next we insert complete sets of Fock states between the factors
$\exp(- \epsilon \beta (H_j - \frac{\mu}{4}N))$.
Each site is either empty or occupied, i.e.
$n_x$ has eigenvalue 0 or 1. In the above representation by Pauli matrices this
corresponds to eigenstates $|0\rangle$ and $|1\rangle$ of $\sigma_l^3$ with
$\sigma_l^3 |0\rangle = -|0\rangle$ and $\sigma_l^3 |1\rangle = |1\rangle$.
One obtains
\begin{eqnarray}
&&\exp(- \epsilon \beta (h_{x,i} - \frac{\mu}{4} n_x -
\frac{\mu}{4} n_{x+\hat{i}})) =
\exp(- \epsilon \beta (1 - \frac{\mu}{4})) \nonumber \\
&&\times \left(\begin{array}{cccc}
\exp(\epsilon \beta (1 - \frac{\mu}{4})) \1 & 0 & 0 & 0 \\
0 & \cosh(\epsilon \beta) \1 & \sinh(\epsilon \beta) \Sig & 0 \\
0 & \sinh(\epsilon \beta) \Sig & \cosh(\epsilon \beta) \1 & 0 \\
0 & 0 & 0 & \exp(-\epsilon \beta (1 - \frac{\mu}{4})) \1 \end{array} \right),
\nonumber \\
\label{matrix}
\end{eqnarray}
where the $4 \times 4$ matrix is in the Fock space basis $|00\rangle$,
$|01\rangle$, $|10\rangle$, $|11\rangle$ of the sites $x$ and $x+\hat{i}$.
$\1$ is the identity and
$\Sig = \sigma_{l+1}^3 \sigma_{l+2}^3 ... \sigma_{m-1}^3$ is a string of Pauli
matrices running over consecutive labels between $l$
and $m$, where $l$ labels $x$ and $m$ labels $x+\hat{i}$. The operators
$\1$ and $\Sig$ act on the remaining
occupation numbers and are diagonal in our basis.
The partition function turns into a path integral over occupation numbers
$n(x,t) = 0,1$  ($t$ labels the time slice)
with periodic boundary conditions in the euclidean time
direction\begin{equation}
Z = \prod_{x,t} \sum_{n(x,t) = 0,1} \exp(- S[n]) \, \mbox{Sign}[n] .
\end{equation}
The Boltzmann factor takes the form
\begin{eqnarray}
&&\exp(- S[n]) = \!\!\! \prod_{x=(2m,n),t=4p}
\exp(- s[n(x,t),n(x+\hat{1},t),n(x,t+1),n(x+\hat{1},t+1)])
\nonumber \\ && \times
\prod_{x=(m,2n),t=4p+1}
\exp(- s[n(x,t),n(x+\hat{2},t),n(x,t+1),n(x+\hat{2},t+1)])
\nonumber \\ && \times
\prod_{x=(2m+1,n),t=4p+2}
\exp(- s[n(x,t),n(x+\hat{1},t),n(x,t+1),n(x+\hat{1},t+1)])
\nonumber \\ && \times
\prod_{x=(m,2n+1),t=4p+3}
\exp(- s[n(x,t),n(x+\hat{2},t),n(x,t+1),n(x+\hat{2},t+1)]), \nonumber \\
\end{eqnarray}
with $s[0,0,0,0] = 0$, $s[1,1,1,1] = 2 \epsilon \beta (1 - \frac{\mu}{4})$,
$s[0,1,0,1] = s[1,0,1,0] = \epsilon \beta (1 - \frac{\mu}{4}) -
\ln \cosh(\epsilon \beta)$,
$s[0,1,1,0] = s[1,0,0,1] = \epsilon \beta (1 - \frac{\mu}{4}) -
\ln \sinh(\epsilon \beta)$.
All other action values are infinite. Note that the occupation numbers
$n(x,t) = 0,1$ are bosonic variables interacting with each other via the
time-like plaquette couplings
$s[n(x,t),n(x+\hat{i},t),n(x,t+1),n(x+\hat{i},t+1)]$.
This structure is identical with the one that occurs in path integral
representations of quantum spin systems \cite{Wie92a}.

In addition to the Boltzmann factor each configuration is
weighted by a sign factor which arises from the strings of
Pauli matrices of eq.(\ref{matrix}). Just as the Boltzmann factor
$\exp(- S[n])$ the sign factor $\mbox{Sign}[n]$ is a product of terms
$\mbox{sign}[n(x,t),n(x+\hat{i},t),n(x,t+1),n(x+\hat{i},t+1)]$ associated with
each plaquette interaction. One has
$\mbox{sign}[0,0,0,0]$ $= \mbox{sign}[1,1,1,1]$ $=
\mbox{sign}[0,1,0,1]$ $= \mbox{sign}[1,0,1,0] = 1$. A nontrivial
sign $\pm 1$ may arise only for plaquette interactions of type $[0,1,1,0]$ and
$[1,0,0,1]$. To compute the sign factors it is convenient to order the
$h_{x,i}$ of each $H_j$ in some (again arbitrary) way.
The factor
$\mbox{sign}[n(x,t),n(x+\hat{i},t),n(x,t+1),n(x+\hat{i},t+1)]$ is the product
of eigenvalues $\pm 1$ of $\sigma_p^3$ for
the sites with label $p$ between $l$ and $m$, i.e. $l+1 \leq p \leq m-1$.
Here $l$ labels the point $x$ and $m$ labels the point $x+\hat{i}$. If the
$h_{y,i}$ to which the point with label $p$ belongs is ordered before
$h_{x,i}$ its $4 \times 4$ matrix of eq.(\ref{matrix}) has already
acted on the site labeled by $p$ and we must take its occupation number at
time $t+1$. If, on the other hand, $h_{y,i}$ is ordered
after $h_{x,i}$ it has not yet acted on this site and we
must take the occupation number at the earlier time $t$. Note that an occupied
site gives a factor 1 while an empty site gives a factor $-1$. Because the
individual $h_{x,i}$ of a given $H_j$ commute with each other
their order does not influence the final result. Moreover, it turns out that
the total sign factor $\mbox{Sign}[n]$ is independent of the chosen ordering
of lattice sites, although the contributions from individual plaquettes
are in general order-dependent.
Hence, any reference to the arbitrary order in the Jordan-Wigner
representation has disappeared from the final expression. This is a very
pleasant surprise.
The sign factor is nonlocal, but it can be computed with an effort
proportional to the lattice size if a convenient ordering is chosen.
More is not needed because the factor is not used
in the updating process. Since it cannot be interpreted as a probability it
is included in the measured observables.
The system without the sign factor is bosonic and interacts
locally. In fact, it corresponds to a quantum spin system with Hamiltonian
$H = \sum_{x,i} (\frac{1}{2} \sigma_x^1 \sigma_{x+\hat{i}}^1 +
\frac{1}{2} \sigma_x^2 \sigma_{x+\hat{i}}^2 + \sigma_x^3 \sigma_{x+\hat{i}}^3)
+
(2 - \frac{\mu}{2}) \sum_x \sigma_x^3$,
which may be viewed as a bosonized version of the original fermionic
model. This type of bosonization up to a nonlocal sign works in any dimension.
In ref.\cite{Amb89} Ambj{\o}rn and Semenoff proceeded in the
opposite direction. They fermionized a quantum spin system without the extra
sign factor and they arrived at $2+1$-dimensional
QED with a Chern-Simons term.
L\"uscher also used a Chern-Simons term in a general construction of
bosonization in $2+1$ dimensions both in the continuum and on the lattice
\cite{Lue89}.

The cluster
algorithm of Evertz, Lana and Marcu \cite{Eve92}, which was originally
constructed for vertex models, works very well also for quantum spin systems
\cite{Wie92b}. Here I describe
a variant of the algorithm suitable for updating the fermionic model. The
algorithm constructs closed loops which are then flipped, i.e.
the occupation numbers of points on the loop are changed from 0 to 1 and vice
versa.
To start a loop one first selects a lattice point $(x,t)$ at random. The
occupation number $n(x,t)$ participates in two plaquette
interactions,
one at euclidean times before and one at euclidean times after $t$. For
$n(x,t) = 1$ we consider the interaction at the later and for $n(x,t) = 0$
we consider the interaction at the earlier time. The corresponding plaquette
configuration
is characterized by the occupation numbers of four lattice
points. One of these points will be the next point on the loop.
For configurations $[0,0,0,0]$ or $[1,1,1,1]$
the next point is with probability
$p = \frac{1}{2}(1 + \exp(- \epsilon \beta))$
the time-like nearest neighbor of $(x,t)$,
and with probability $1 - p$ the next-to-nearest
(diagonal) neighbor of $(x,t)$ on the plaquette.
For configurations $[0,1,0,1]$ or $[1,0,1,0]$
the next point on the loop is
with probability $p/\cosh(\epsilon \beta)$ the time-like nearest
neighbor, and with probability $1 - p/\cosh(\epsilon \beta)$ the
space-like nearest neighbor of $(x,t)$. Finally, for configurations
$[0,1,1,0]$ or $[1,0,0,1]$ the next point is with probability
$(1 - p)/\sinh(\epsilon \beta)$ the diagonal neighbor, and with
probability $1 - (1-p)/\sinh(\epsilon \beta)$ the space-like
nearest neighbor of $(x,t)$. Once the next point on the loop is
determined the process is repeated until the loop closes. The above
probabilities are arranged such that the algorithm obeys detailed balance.
The part of the action proportional to $1 - \frac{\mu}{4}$
is taken into account by a global Metropolis step. For this purpose
each loop ${\cal C}$ is characterized by a winding number $W({\cal C})$,
which counts how often the loop winds
around the lattice in the euclidean time direction. The winding number is
related to the total occupation number of the loop
$W({\cal C}) = \frac{1}{4M} \sum_{(x,t) \in {\cal C}} (2n(x,t) - 1)$
where $4M$ is the number of euclidean time slices.
The action associates a Boltzmann factor
$\exp(- \beta (2 - \frac{\mu}{2}) W({\cal C}))$ with each loop.
When the loop is flipped
its winding number changes sign. In the Metropolis step
the loop is flipped with probability
$p = \mbox{min}(1,\exp(\beta (4 - \mu) W({\cal C})))$.

I have applied the algorithm in a single cluster version for various values
of $\beta$ and $\mu$ at fixed lattice spacing $\epsilon \beta = 1/16$.
For equilibration 100000 loop clusters have been updated,
followed by 100000 measurements each separated by 10 loop updates. The
Monte-Carlo (MC) results are compared to the exact results in table 1; both
agree within error bars. Note that $\langle n_x \rangle =
\langle n_x \mbox{Sign}[n] \rangle_b/\langle
\mbox{Sign}[n] \rangle_b$ where $b$ refers to
the simulated bosonic ensemble. One can see that the minus-sign problem
becomes more severe when the temperature is lowered or when the chemical
potential is increased. However, only at $\beta = 1$, $\mu = 4$ there is
a real problem, because only then $\langle \mbox{Sign}[n] \rangle_b$
is consistent with zero. A detailed analysis of the efficiency of the algorithm
will be presented elsewhere.
\begin{table} \begin{center}
\caption{Results of the numerical simulations.}
\vspace{0.2cm}
\begin{tabular}{|c|c|c|c|c|c|c|} \hline
$\beta$ & $\mu$ & $L$ & $4M$ & $\langle \mbox{Sign}[n] \rangle_b$ &
$\langle n_x \rangle_{\mbox{MC}}$ & $\langle n_x \rangle_{\mbox{exact}}$
\\ \hline
0.25 & 0 & 8 & 16 & 0.976(1) & 0.2798(3) & 0.27961 \\ \hline
0.25 & 1 & 8 & 16 & 0.970(1) & 0.3304(4) & 0.32989 \\ \hline
0.25 & 2 & 8 & 16 & 0.964(1) & 0.3843(5) & 0.38417 \\ \hline
0.25 & 4 & 8 & 16 & 0.963(1) & 0.4999(7) & 0.50000 \\ \hline
0.50 & 0 & 8 & 32 & 0.868(2) & 0.1555(5) & 0.15655 \\ \hline
0.50 & 1 & 8 & 32 & 0.779(2) & 0.223(1)  & 0.22310 \\ \hline
0.50 & 2 & 8 & 32 & 0.686(2) & 0.306(1)  & 0.30493 \\ \hline
0.50 & 4 & 8 & 32 & 0.587(2) & 0.500(3)  & 0.50000 \\ \hline
1.00 & 0 & 8 & 64 & 0.747(2) & 0.067(1)  & 0.06853 \\ \hline
1.00 & 1 & 8 & 64 & 0.322(3) & 0.137(2)  & 0.13509 \\ \hline
1.00 & 2 & 8 & 64 & 0.052(3) & 0.23(2)   & 0.23209 \\ \hline
1.00 & 4 & 8 & 64 &-0.001(3) & 0.5(1.7)  & 0.50000 \\ \hline
\end{tabular} \end{center} \end{table}

The purpose of the present paper was to describe the bosonization scheme for
lattice fermions, and to apply it to numerical simulations of a simple model.
Of course, the idea is to apply the method to models of physical interest.
In condensed matter
physics one can attack the Hubbard model or other
models relevant for high $T_c$ superconductivity. In relativistic lattice
field theories one may first study free Wilson or staggered fermions.
The simplest nontrivial particle physics
application is perhaps to the purely fermionic Gross-Neveu model. In
general, however, the fermions are coupled to bosonic fields as well.
For example,
the Yukawa coupling to a scalar field corresponds to a space-time dependent
chemical potential and is easy to incorporate.
In QCD one wants to study the coupling of quarks to $SU(3)$ gauge fields,
which should be possible along similar lines.
In all cases the minus-sign problem might raise its ugly
head once one enters a physically interesting regime. Only a detailed
investigation of the various models of interest can show if this is the
case or not. Work in some of these directions is in progress.

I like to thank P. Wiese for a very interesting discussion.


\end{document}